\documentclass[11pt]{article}
\usepackage{graphicx}
\usepackage{amssymb}

\def\beqa{\begin{eqnarray}}
\def\eeqa{\end{eqnarray}}
\def\beq{\begin{equation}}
\def\eeq{\end{equation}}

\begin{document}

\title{Improved tests on the relationship between the kinetic energy of  galaxies
and the mass of their central black holes}
\author{A. Feoli\footnote{E-mail: feoli@unisannio.it} and D. Mele\\ Dipartimento di Ingegneria, Universit\`{a} del
            Sannio, \\Corso Garibaldi n. 107, Palazzo Bosco Lucarelli  \\ I-82100 - Benevento, Italy.}
\date{}

 \maketitle

\begin{abstract}We support, with new fitting instruments and the
analysis of more recent experimental data, the proposal of a
relationship between the mass of a Supermassive Black Hole (SMBH)
and the kinetic energy of random motions in the host elliptical
galaxy. The first results obtained in a previous paper with 13
elliptical galaxies are now confirmed by the new data and an
enlarged sample. We find $M_{BH} \propto (M_{G}
\sigma^{2}/c^2)^\beta$ with $0.8 \leq \beta \leq 1$ depending on
the different fitting methods and samples used. The meaningful
case $\beta = 1$ is carefully analyzed. Furthermore, we test the
robustness of our relationship including in the sample also
lenticular and spiral galaxies and we show that the result does
not change. Finally we find a stronger correlation between the
mass of the galaxy and the corresponding velocity dispersion that
allows to connect our relationship to the $M_{BH} \propto
\sigma^\alpha$ law. With respect to this law, our relationship has
 the advantage to have a smaller scatter.
\end{abstract}

keywords: black hole physics --  galaxies: kinematics and dynamics

\section{Introduction}

In the last ten years the existence of Supermassive Black Holes in
the central part of an increasing number of galaxies has become a
stronger and stronger evidence. In order to understand the
formation process and the evolution of those black holes, it is
important to relate them to the properties of the host galaxy. It
was shown that the existence of a central SMBH
 affects the dynamics not only in the core of the host galaxy but
 even in a region far from the hole.
 Several relationships have been recently proposed between the mass of
a central SMBH and the velocity dispersion  \cite{fer,geb,tre},
the bulge luminosity or mass
 \cite{kor,mag,mar,mer,pin,rich,van}  or the dark matter halo \cite{fer2},
 of the corresponding galaxy. Among them,
the relationship with the smallest scatter is $$M_{BH} \propto
\sigma^{\alpha} \eqno(1)$$ where $3.75 < \alpha < 5.3$. The
different values of $\alpha$, found by several authors using
different samples and different fitting methods, are well
discussed in the paper of Tremaine et al. \cite{tre} that obtain
$$Log M_{BH} = (4.02 \pm 0.32) Log (\sigma/200) + (8.13 \pm 0.06)
\eqno(2)$$ where (and also afterwards in the rest of our paper)
$M_{BH}$ is expressed in solar masses, $\sigma$ in $km/s$ and
logarithms are base 10.

From the theoretical point of view there are still different
interpretations of these results \cite{bur,dok,hae,ada} and in
order to give a contribution to this open debate, in a previous
paper \cite{feo}, we proposed to study a relationship between the
kinetic energy of random motions of elliptical galaxies and the
rest energy of their central Supermassive Black Hole. We used a
sample formed by the intersection between the set of galaxies
studied by Tremaine et al. \cite{tre} and the kinematical data
extracted by \cite{blf,busl,cur} that refer to the effective
semimajor axis of each galaxy $A_e$ (related to the effective
radius $R_e$ by the formula
$$\frac{R_e}{A_e}= 1 - \frac{\epsilon}{2} \eqno(3)$$ where
$\epsilon$ is the ellipticity). The model used to compute the
kinematical parameters is  very simple. Busarello, Longo and Feoli
\cite{blf} assumed that an elliptical galaxy
\begin{enumerate}
\item has a spheroidal symmetry, \item follows the de Vaucouleurs
$r^{1/4}$ law, \item its rotation axis is perpendicular to the
line of sight, \item its stars have a rotation velocity with
cylindrical symmetry and \item its  velocity dispersion tensor is
isotropic and has spherical symmetry. \end{enumerate} The details
of the procedure used to compute velocity dispersions, rotation
velocities and the masses of the 13 elliptical galaxies included
in the so called sample A (see Table 1) were fully explained in
\cite{blf,busl} and summarized in our previous paper \cite{feo}
(hereafter AFDM).

Our first results were very encouraging because we found, taking
into account errors in both variables and using an iterative
procedure due to Orear \cite{or}:
$$Log(M_{BH}) = (0.88 \pm 0.06) Log \frac{M_G
\sigma^{2}}{c^2} + (4.30 \pm 0.46) \eqno(4)$$ with (see Appendix
for definitions) $\chi^2_r =2.2$ and with a linear correlation
coefficient $r = 0.833$. If the Akritas and Bershadi method
\cite{akr} is used, the slope of the relationship is even closer
to unity:
$$Log(M_{BH}) = (0.98
\pm 0.09) Log \frac{M_G \sigma^{2}}{c^2} + (3.8 \pm 0.4)
\eqno(5)$$ with $\chi^2_r = 1.9$.

 Even better results can be obtained using a reduced sample
(hereafter SAMPLE B) of galaxies if we eliminate the two
ellipticals with the largest residuals N821 and N4697. The
relationships obtained applying the fitting procedure to the
remaining 11 galaxies are listed in AFDM. The satisfying results
are the increase of the correlation coefficient, the decrease of
$\chi^2$, and a slope closer to unity.

In order to understand better the origin of this relationship, we
have performed in this paper four new tests:
\begin{enumerate}
\item We used the well known iterative procedure of FITEXY routine
(we had never used before) to analyze again the same samples A and
B of the previous paper AFDM in the case of errors in both
variables;

\item The simplest hypothesis of a linear relationship was tested
using the exact "least-squares" fitting method;

\item The kinematical data used in the previous paper were
extracted by old sources so we have performed a check of the
relationship using the data recently published by H\"{a}ring and
Rix \cite{pin}.

\item In AFDM we have considered only elliptical galaxies.
 In this paper we  tested the robustness of the relationship, including in
the statistical analysis also the lenticular and spiral galaxies
of the sample published by H\"{a}ring and Rix  \cite{pin}.
\end{enumerate}
There is another remarkable novelty  contained in this paper.
Actually in AFDM we found a very poor correlation between the mass
of the galaxy and its velocity dispersion. On the contrary with
the sample of H\"{a}ring and Rix \cite{pin} this relationship is
stronger and we will show that if
$$M_{BH} \propto \sigma^{\alpha} \eqno(6)
$$ and $$M_{BH} c^2 \propto (M_G \sigma^{2})^\beta \eqno(7)$$ then (as we
expected) a relation $$M_G \propto \sigma^\gamma \eqno(8)$$ is
such that
$$\gamma = \frac{\alpha}{\beta} - 2 \eqno(9)$$
Furthermore, using the new data, we will show that the equation
(7) has a  $\chi^2$ smaller than the relation (6).
\section{The samples}

\begin{table}

\begin{center}
\title{The old data.}
\end{center}

\centering
\begin{tabular}{c c c c c c}
\hline\hline
(1) & (2)& (3)& (4)& (5) & (6)  \\
{Galaxy} & $M_{BH}$ & $\sigma$ & V & $M_G$ & $\frac{\Delta M_G}{M_G}$\\
{} & {($M_{\odot}$)} & {(km/s)} & {(km/s)} & {($M_{\odot}$)} & {}\\
\hline
N221  &    $2.5 \times 10^6$   &   60  &   37  &   $4.0\times 10^{8}$  &   0.07    \\
  N821  &     $3.7 \times 10^7$   &   180 &   117 &   $1.4 \times 10^{11}$  &   0.04    \\
  N2778 &     $1.4 \times 10^7$   &   140 &   96  &   $4.0 \times 10^{10}$  &   0.16    \\
  N3379 &     $1.0 \times 10^8$   &   193 &   53  &   $3.0 \times 10^{10}$  &   0.12    \\
  N4291 &     $3.1 \times 10^8$   &   250 &   76  &   $9.0 \times 10^{10}$  &   0.05    \\
  N4473 &     $1.1 \times 10^8$   &   191 &   62  &   $7.0 \times 10^{10}$  &   0.16    \\
  N4486 &     $3.0 \times 10^9$   &   269 &   20  &   $3.5 \times 10^{11}$  &   0.10    \\
  N4564 &     $5.6 \times 10^7$   &   125 &   143 &   $5.0 \times 10^{10}$  &   0.16    \\
  N4649 &     $2.0 \times 10^9$   &   224 &   56  &   $2.2 \times 10^{11}$  &   0.51    \\
  N4697 &     $1.7 \times 10^8$   &   177 &   151 &   $2.9 \times 10^{11}$  &   0.19    \\
  N4742 &     $1.4 \times 10^7$   &   81  &   91  &   $1.0 \times 10^{10}$  &   0.17    \\
  N5845 &     $2.4 \times 10^8$   &   236 &   73  &   $2.0 \times 10^{10}$  &   0.16    \\
  N6251 &     $5.3 \times 10^8$   &   288 &   54  &   $2.3 \times 10^{11}$  &   0.12    \\
  IC1459&     $2.5 \times 10^9$   &   282 &   383 &   $3.5 \times 10^{11}$  &   0.03    \\
\hline
\end{tabular}
\caption{We list in column (1) the name of the galaxy, in column
(2) the black hole masses taken from Tremaine et al. \cite{tre}.
The velocity dispersions (column 3) and the rotation velocities
(column 4) are taken from Busarello, Feoli and Longo \cite{blf},
the masses in column (5) from Curir  et al.\cite{cur} and the
corresponding relative errors in column (6) from Busarello and
Longo \cite{busl}. The sources of the experimental data for each
galaxy are listed in the above mentioned papers.}
\end{table}

 We decide to adopt
in the sample A and B  the kinematical parameters computed by
Busarello et al. and published for $\sigma$ and $V$ of 54
elliptical galaxies in \cite{blf} and for the mass and the
specific angular momentum in \cite{busl,cur}. These have the
advantage to be treated with the same method and the same fitting
procedure; furthermore they are all referred to the effective
semimajor axis,
 and are corrected for one projection effect: the integration of the
  light along the line of sight. All the masses are computed
  using the same method that is the Jeans equation describing the equilibrium of a spheroidally symmetric
system having an isotropic velocity dispersion tensor.
 The intersection between the set of elliptical galaxies studied
in \cite{blf,busl,cur} and the set of SMBH masses in Table 1 in
the paper of Tremaine et al. \cite{tre} leads to 13 galaxies (the
elliptical N221 was eliminated because it has a very low velocity
dispersion and a mass two orders of magnitude less than all the
others) that are listed in Table 1 where $V$ is the luminosity
weighted mean rotation velocity inside $A_e$ (related to the
rotational kinetic energy by $T_V = M V^2/2$) and  $$ \sigma^2
\equiv <\sigma^2_{yy}> = \frac{\int \sigma^2_{yy} (r) \rho d^3 x
}{\int \rho d^3 x} = \frac{2T_\sigma}{3M} \eqno(10)$$ is the
luminosity weighted mean of the line of sight component, of the
velocity dispersion tensor, assuming that the mass-to-light ratio
is constant inside $A_e$ and that the tensor is isotropic
($T_\sigma$ is the corresponding kinetic energy).

 The
sample A is formed by all these 13 galaxies while, as in AFDM, we
will denote with sample B the set obtained from A eliminating the
galaxies: N821 and N4697.

 In order to check our new
relationship with a set of more recent data, we extract from the
Table 1 of the paper of H\"{a}ring and Rix \cite{pin} the values
listed in the first part of our Table 2, including 18 elliptical
galaxies (all the ellipticals except N221) that form our sample C.
If we add also the remaining eleven non elliptical galaxies of
H\"{a}ring and Rix \cite{pin}, we obtain our largest sample D.
Finally we will study with the new data of Table 2 the same set of
galaxies of the sample A and B and we will denote the
corresponding samples with $A'$ and $B'$. Of course the methods
used to compute velocity dispersions and Bulge masses by
H\"{a}ring and Rix \cite{pin} are often different from Busarello
et al. and the data do not refer to the semimajor axis of each
galaxy but can extend also to a bulge region of radius $3R_e$.
This justifies that sometime there are significant differences in
the values of the parameters found in the two tables for the same
galaxies. For the masses of Black Holes we have preferred to use
all the set in \cite{tre} even if, for a few galaxies, now some
updated values exist, measured by other authors.
\begin{table}

\begin{center}
\title{The new data.}
\end{center}
\centering
\begin{tabular}{c c c c c c}
\hline\hline
(1) & (2)& (3)& (4)& (5) & (6) \\
{Galaxy} & {Type} & {$M_{BH}$} & {$\Delta M_{BH}$} & {$\sigma$} & {$M_G$} \\
{} & {} & {($M_{\odot}$)} & {($M_{\odot}$)} & {(km/s)} & {($M_{\odot}$)} \\
\hline
 N821   &   E4  &   $3.7 \times 10^7$   &   $2.4 \times 10^7$   &   209 &   $1.3 \times 10^{11}$  \\
 N2778  &   E2  &   $1.4 \times 10^7$   &   $0.9 \times 10^7$   &   175 &   $7.6 \times 10^{10}$  \\
 N3377  &   E5  &   $1.0 \times 10^8$   &   $0.9 \times 10^8$   &   145 &   $1.6 \times 10^{10}$  \\
 N3379  &   E1  &   $1.0 \times 10^8$   &   $0.6 \times 10^8$   &   206 &   $6.8 \times 10^{10}$  \\
 N3608  &   E2  &   $1.9 \times 10^8$   &   $1.0 \times 10^8$   &   182 &   $9.7 \times 10^{10}$  \\
 N4261  &   E2  &   $5.2 \times 10^8$   &   $1.1 \times 10^8$   &   315 &   $3.6 \times 10^{11}$  \\
 N4291  &   E2  &   $3.1 \times 10^8$   &   $2.3 \times 10^8$   &   242 &   $1.3 \times 10^{11}$  \\
 N4374  &   E1  &   $4.3 \times 10^8$   &   $3.2 \times 10^8$   &   296 &   $3.6 \times 10^{11}$  \\
 N4473  &   E5  &   $1.1 \times 10^8$   &   $7.9 \times 10^7$   &   190 &   $9.2 \times 10^{10}$  \\
 N4486  &   E0  &   $3.0 \times 10^9$   &   $1.0 \times 10^9$   &   375 &   $6.0 \times 10^{11}$  \\
 N4564  &   E3  &   $5.6 \times 10^7$   &   $0.8 \times 10^7$   &   162 &   $4.4 \times 10^{10}$  \\
 N4649  &   E1  &   $2.0 \times 10^9$   &   $0.6 \times 10^9$   &   375 &   $4.9 \times 10^{11}$  \\
 N4697  &   E4  &   $1.7 \times 10^8$   &   $0.2 \times 10^8$   &   177 &   $1.1 \times 10^{11}$  \\
 N4742  &   E4  &   $1.4 \times 10^7$   &   $0.5 \times 10^7$   &   90  &   $6.2 \times 10^{9}$  \\
 N5845  &   E3  &   $2.4 \times 10^8$   &   $1.4 \times 10^8$   &   234 &   $3.7 \times 10^{10}$  \\
 N6251  &   E2  &   $5.3 \times 10^8$   &   $1.8 \times 10^8$   &   290 &   $5.6 \times 10^{11}$  \\
 N7052  &   E4  &   $3.3 \times 10^8$   &   $2.3 \times 10^8$   &   266 &   $2.9 \times 10^{11}$  \\
 IC1459 &   E3  &   $2.5 \times 10^9$   &   $0.5 \times 10^9$   &   323 &   $2.9 \times 10^{11}$  \\
\hline
 N224   &   Sb  &   $4.5 \times 10^7$   &   $4.0 \times 10^7$   &   160 &   $3.7 \times 10^{10}$  \\
 N1023  &   SB0 &   $4.4 \times 10^7$   &   $0.5 \times 10^7$   &   205 &   $6.9 \times 10^{10}$  \\
 N1068  &   Sb  &   $1.5 \times 10^7$   &   $1.5 \times 10^7$   &   151 &   $2.3 \times 10^{10}$  \\
 N3115  &   S0  &   $1.0 \times 10^9$   &   $1.0 \times 10^9$   &   230 &   $1.2 \times 10^{11}$  \\
 N3245  &   S0  &   $2.1 \times 10^8$   &   $0.5 \times 10^8$   &   205 &   $6.8 \times 10^{10}$  \\
 N3384  &   S0  &   $1.6 \times 10^7$   &   $0.2 \times 10^7$   &   143 &   $2.0 \times 10^{10}$  \\
 N4342  &   S0  &   $3.0 \times 10^8$   &   $1.7 \times 10^8$   &   225 &   $1.2 \times 10^{10}$  \\
 N4594  &   Sa  &   $7.8 \times 10^7$   &   $4.2 \times 10^7$   &   240 &   $2.7 \times 10^{11}$  \\
 N7332  &   S0  &   $1.3 \times 10^7$   &   $0.6 \times 10^7$   &   122 &   $1.5 \times 10^{10}$  \\
 N7457  &   S0  &   $3.5 \times 10^6$   &   $1.4 \times 10^6$   &    67 &   $7.0 \times 10^{09}$  \\
 Milky Way &  SBbc   &   $1.8 \times 10^6$   &   $0.4 \times 10^6$   &   75 &   $1.1 \times 10^{10}$  \\
\hline
\end{tabular}
\caption{We list in column (1) the name of the galaxy, in column
(2) the morphological type from de Vaucouleurs et al.\cite{vac},
in column (3) the black hole masses and in column (4) the
corresponding errors both taken from Tremaine et al. \cite{tre}
(except for N4374, N4594 and N7332 taken from \cite{pin}), in
column (5) the stellar velocity dispersions and in column (6) the
galaxy masses both taken from H\"{a}ring and Rix \cite{pin}. The
sources of the experimental data for each galaxy are listed in the
above cited papers.}
\end{table}

\section{The new results}

We start using Fitexy Routine to take into account errors on both
variables. We apply it to our six samples of galaxies and the
results are reported in Table 3.
\begin{table}

\centering
\begin{tabular}{c c c c c c}
\hline\hline
(1) & (2)& (3)& (4)& (5) & (6) \\
{Sample} & {N} & {$m\pm\Delta m$} & {$b\pm\Delta b$} & {$\chi^2$}
& {r}\\
\hline
$A$  &   13  &   $1.04\pm0.12$ &   $3.53\pm0.58$ &   23.52   &   0.83    \\
$B$  &   11  &   $0.99\pm0.11$ &   $3.89\pm0.54$ &   11.53   &   0.91    \\
\hline
$A^{\prime}$ &   13  &   $0.86\pm0.08$ &   $4.23\pm0.42$ &   26.19   &   0.84    \\
$B^{\prime}$ &   11  &   $0.84\pm0.08$ &   $4.36\pm0.41$ &   20.36   &   0.88    \\
$C$  &   18  &   $0.80\pm0.07$ &   $4.48\pm0.36$ &   32.47   &   0.82    \\
$D$  &   29  &   $0.90\pm0.05$ &   $4.03\pm0.24$ &   55.02   &   0.87    \\
\hline
\end{tabular}
\caption{We denote in column (1) the sample, in column (2) the
corresponding number of galaxies. We find with Fitexy the best fit
of the relationship $\log(M_{BH}) = m \cdot
\log(M_{G}\sigma^2/c^2) + b$. The results for $m$ and $b$ are in
column (3) and (4) and the corresponding $\chi^2$ in column (5).
Finally the linear correlation coefficient is in column (6).}
\end{table}
 We considered for the sample A
and B the worst scenario of a relative error $20\%$ on $\sigma$ as
we estimated in AFDM from the discussion on the possible sources
of error (anisotropy, triaxiality, fitting procedure, inclination,
etc.), while the relative errors on the masses $M_G$ computed in
\cite{busl} are listed in Table 1. On the contrary, for the data
of H\"{a}ring and Rix \cite{pin} listed in table 2 (used in the
samples $A^\prime$, $B^\prime$, $C$ and $D$), we consider a
relative error on velocity dispersion of $5\%$ and a
$\Delta(LogM_{G})=0.18$.

The trend for the ellipticals is that the old data (samples A and
B) give slopes closer to unity while the new data have slopes near
to $0.8$. The best fit line
$$Log(M_{BH}) = (0.80
\pm 0.07) Log \frac{M_G \sigma^{2}}{c^2} + (4.5 \pm 0.4)
\eqno(11)$$ of the largest sample of ellipticals C is shown in
Fig. 1. In order to explain these results, we must take into
account that the new data are no more averaged on the semimajor
axis and have a different distribution of errors, but the reason
could be more simply due to the different fitting methods because
the Orear iterative procedure gave with the old data  similar
results (see eq. 4). It is not surprising that two iterative
procedures can give different outputs because, having different
starting points, they can stop in different local minima of
$\chi^2$. Certainly an important role is also played by the
relationship (8) between the masses of the galaxies and the
corresponding velocity dispersions. In fact in AFDM the
correlation coefficients of this relation were very low ($r_A=
0.678$ and $r_B = 0.755$) involving that the two parameters could
be considered independent from each other. Now these coefficients
are very high ($r_{A'}= 0.916, r_{B'}= 0.932$ and $r_C = 0.923$)
so the mass depends on the velocity dispersion,
 the relationship (7) can be expressed in terms of only one
parameter, and the laws $M_{BH} \propto \sigma^{\alpha}$ and
$M_{BH} \propto M_G^{\delta}$ have a common origin. To this aim we
have listed in Table 4 the slopes of the relationships (6), (7),
and (8)  between the parameters, and we have checked if
$(\alpha/\beta) - 2 = \gamma$. It is very clear and remarkable,
comparing the last two columns, the agreement between the slopes
of the first two relationships with the third one.
\begin{table}

\centering
\begin{tabular}{c c c c c}
\hline\hline
(1) & (2)& (3)& (4)& (5)\\
{Sample}&{$\alpha\pm\Delta {\alpha}$} &{$\beta\pm\Delta {\beta}$}&
{$\gamma\pm\Delta
{\gamma}$}& {$\frac{\alpha}{\beta}-2$}\\
{} & {$(\chi^2)$} & {$(\chi^2)$} & {$(\chi^2$)}& {}\\
\hline
$A^{\prime}$  &   $4.24\pm0.28$ &   $0.86\pm0.08$ &   $3.06\pm0.33$ &   2.93    \\
    & (32.38) & (26.19)  & (16.13) & \\
$B^{\prime}$  &   $4.32\pm0.28$ &   $0.84\pm0.08$ &   $3.14\pm0.34$ &   3.14    \\
    & (21.67) & (20.36)  & (13.07) & \\
$C$   &   $4.02\pm0.26$ &   $0.80\pm0.07$ &   $3.23\pm0.30$ &   3.02    \\
    & (42.08) & (32.47)  & (20.07) & \\
$D$   &   $4.44\pm0.17$ &   $0.90\pm0.05$ &   $2.86\pm0.19$ &   2.93    \\
    & (89.99) & (55.02)  & (59.94) & \\
\hline
\end{tabular}
\caption{We denote in column (1) the sample, in column (2) the
best fit for the relationship $M_{BH}\propto\sigma^{\alpha}$, in
column (3) for $M_{BH}\propto(M_{G}\sigma^2/c^2)^{\beta}$, and in
column (4) for $M_{G}\propto\sigma^{\gamma}$ (in parentheses we
list the corresponding $\chi^2$). Finally, in column (5), we
compute $\frac{\alpha}{\beta}$ - 2 to compare with $\gamma$.}
\end{table}

It is very important also to stress in Table 4 that the values of
$\chi^2$ of the relationship involving the kinetic energy (column
3) are lower than the more famous first relationship (6) (column
2) that for the sample C can be written:
$$Log M_{BH} = (4.02 \pm
0.26) Log (\sigma/200) + (8.23 \pm 0.05) \eqno(12)$$
 and it is (as we expected) in full agreement with the result of Tremaine et al (2).
Our relationship (7) has a better $\chi^2$ even with respect to
the $M_{BH}$ vs $M_G$ law. Using the data in Table 2 for the
largest sample D, we obtain
$$ Log M_{BH} = (1.49 \pm 0.11) Log M_G - (8.15 \pm 1.20) \eqno(13)$$ with a
$\chi^2 = 63$ compared with $\chi^2 = 55$ of our law.

Furthermore (unlike AFDM) we find that the relation between the
mass of the elliptical galaxies and the velocity dispersion
(column 4 of Table 4) has the smallest scatter. From the
dependence of the mass-to-light ratios on the luminosity $M_G/L
\propto L^{0.25}$ and from the Faber - Jackson relation $L \propto
\sigma^4$, Ferrarese and Merritt \cite{fer} infer $M_G \propto
\sigma^5$ for early type galaxies. Then, following the reasoning
of Burkert and Silk \cite{bur}, we could expect from the
application of the Virial theorem $(M_G \propto R_e \sigma^2)$ and
from the results of the Fundamental Plane of elliptical galaxies
$(R_e \propto \sigma^2)$ that $M_G \propto \sigma^4$, but the
best-fit of sample C  gives:
$$Log M_{G} = (3.2 \pm 0.3) Log (\sigma/200) + (10.93 \pm 0.05)
\eqno(14)$$ This result cannot be explained with the Faber -
Jackson relation while it could be in agreement with the $L\propto
\sigma^2$ or $L \propto \sigma^{2.5}$ obtained for faint
ellipticals \cite{dav,mat}. The very interesting fit of equation
(14) is shown in Fig. 2 where it is evident that the galaxy N5845
(denoted by a triangle) has the largest residual. Eliminating this
galaxy, the best fit line (14) does not change while the linear
correlation coefficient increases becoming $r_C = 0.958$ and the
value of $\chi^2$ decreases from 20.07 to 10.55 resulting in a
very strong correlation.

While a coefficient less than one as $0.8$ is more difficult to be
interpreted, it would be more interesting, from the theoretical
point of view, the meaningful possibility of an exact linear
relationship between the mass of a SMBH and the kinetic energy of
random motions of the host galaxy. It is very simple to test this
hypothesis using the exact formulas (without any approximation or
iterative procedure) of the standard "least-squares method" that
we have reported in the appendix.
\begin{table}
\centering
\begin{tabular}{c c c c c}
\hline\hline
(1) & (2)& (3)& (4)& (5)\\
{Sample} &{N} &{m} &{$b \pm \Delta b $} &{$\chi^2$}\\
\hline
$A$   &   13  &   1   &   $3.73\pm0.08$ &   23.67   \\
$B$   &   11  &   1   &   $3.86\pm0.09$ &   11.54   \\
\hline
$A^{\prime}$  &   13  &   1   &   $3.56\pm0.08$ &   28.31   \\
$B^{\prime}$  &   11  &   1   &   $3.59\pm0.09$ &   23.19   \\
$C$   &   18  &   1   &   $3.52\pm0.07$ &   37.91   \\
$D$   &   29  &   1   &   $3.58\pm0.06$ &   58.20   \\
\hline
\end{tabular}
\caption{We denote in column (1) the sample, in column (2) the
corresponding number of galaxies, in column (3) we fix the
coefficient $m = 1$ and we find with Fitexy the best fit of the
relationship $\log(M_{BH}) = m \cdot \log(M_{G}\sigma^2/c^2) + b$.
The result for $b$ is in column (4) and the corresponding $\chi^2$
in column (5).}
\end{table}
The results are listed in Table 5 and we argue that for the sample
B the hypothesis is certainly right, and for the samples A and
$B'$ it can be acceptable. Only the samples $A'$ and C have a too
high $\chi^2$.  The linear fit of sample C is shown in fig. 1 with
a dashed line and involves $M_{BH} c^2 \propto M_G \sigma^2$.

The fourth test for our relationship is the inclusion of
lenticular and spiral galaxies in the statistical analysis. Our
sample D is now composed by all the 29 galaxies listed in Table 2.
In Fig. 3 we show with a dashed line the best fit (11) for the 18
ellipticals (denoted with boxes), with the dot - dashed line the
best fit for the remaining 11 galaxies (denoted with circles):
$$Log(M_{BH}) = (1.14 \pm 0.15)  Log\frac{M_G \sigma^{2}}{c^2} + (3.14 \pm 0.57)  \eqno(15)$$
and with a solid line the best fit for the whole sample D
$$Log(M_{BH}) = (0.90 \pm 0.05)  Log\frac{M_G \sigma^{2}}{c^2} + (4.03 \pm 0.24)  \eqno(16)$$
We argue that the contribution of lenticular and spiral galaxies
increases the slope that becomes closer to unity.

\section{Conclusions}
From the first samples of data (A and B) we were induced in a
previous paper \cite{feo} to consider a relationship between the
masses of SMBHs and the kinetic energy of random motions in
elliptical galaxies. This  suggestion is now confirmed first by
reanalyzing the old data using the Fitexy Routine and then
comparing the results with the corresponding fit of a new set of
kinematical data extracted from the paper of H\"{a}ring and Rix
\cite{pin}. In the light of Fig. 3 we  show that the relationship
works well also for lenticular and spiral galaxies. The remarkable
result is that the relationship we proposed (7), has now a smaller
scatter with respect to the old equation (6) and the $M_{BH}$ vs
$M_G$ law. Furthermore we have shown that, using the new data, all
these relationships can be strictly connected because (unlike
AFDM) the mass of the galaxy strongly depends on the velocity
dispersion: the linear correlation coefficient of this
relationship (8) is very high and the $\chi^2$ very low. Finally
our result (14) with $\gamma = 3.2$ is surprisingly smaller than
the ones expected of $\gamma = 4$ or $\gamma = 5$.

 For the sample A and B, given the assumptions
of the model of Busarello, Longo and Feoli \cite{blf}, it is clear
that triaxiality, anisotropy of the velocity field, inclination,
deviations from the $r^{1/4}$ law, are all sources of possible
errors that could affect the derived kinematical parameters. For
the remaining samples we must consider that the masses are not
computed with the same method and often refer to a bulge region
that ends to $r_{MAX} = 3 R_e$. At the same time all the velocity
dispersions are no longer averaged until the effective semimajor
axis of each galaxy. So the nonhomogeneity of the measurements and
the differences in the part of the galaxy that each single measure
now considers, can affect also the results obtained with the new
set of data. Of course a rigorous analysis will be possible only
when the observational uncertainties in all quantities are reduced
and the sample is further extended. However, though with the
caution due to all these possible error sources, the study of the
relationship we propose between the rest energy of a SMBH and the
kinetic energy of random motions of the host galaxy appears to be
very useful for a more complete understanding of the formation and
evolution of SMBH not only in elliptical galaxies.

\bigskip
{\bf Acknowledgements}
\bigskip

 We are grateful to Gaetano Scarpetta,
Antonio D'Onofrio and Nicola De Cesare for useful comments and
Franco Caprio for helpful suggestions about computer programs. The
research was partially supported by FAR fund  of the University of
Sannio.

\section*{Appendix}

We have tried to maximize the errors so we have always chosen in
the Table 1 of Tremaine et al.\cite{tre} the maximal difference
between the measure of the black hole mass and its high or  low
values in parenthesis. So, for example, for NGC 3379 the measure
$M_{BH} = 1.0 \times 10^8 (0.5, 1.6)$ becomes in our statistical
elaboration (see column 2 of Table 2) $M_{BH} = (1.0 \pm 0.6)
\times 10^8$.

Accordingly, the formula used in this paper to estimate all the
maximal errors in the functions F of the parameters $(a,b,c,....)$
is
$$\Delta F(a,b,c) = \left| \frac{\partial F}{\partial a}\right|
\Delta a + \left|\frac{\partial F}{\partial b}\right| \Delta b +
\left|\frac{\partial F}{\partial c}\right| \Delta c \eqno(A1)$$
Furthermore the reduced $\chi^2$ is defined as:
$$\chi^2_r = \frac{\chi^2}{N-1} = \frac{1}{N-1} \sum_{i=1}^{N} \frac{(y_i
-b-m x_i)^2}{(\Delta y_i)^2 + m^2 (\Delta x_i)^2}\eqno(A2)$$ for a
relation of the form $y = b + m x$.

In the case $m = 1$ we have used  the exact formulas of the least
squares method:
$$b = \frac{\sum_{i=1}^{N} \left(\frac{y_i -x_i}{(\Delta x_i)^2 + (\Delta
y_i)^2}\right)}{\sum_{i=1}^{N} \left(\frac{1}{(\Delta x_i)^2 +
(\Delta y_i)^2} \right)} \eqno(A3)$$
$$(\Delta b)^2 = \frac{1}{\sum_{i=1}^{N} \left(\frac{1}{(\Delta x_i)^2 + (\Delta
y_i)^2}\right)} \eqno(A4)$$ and the results are reported in Table
5.
\newpage

\begin{figure*}
\centering
\resizebox{\hsize}{!}{\includegraphics{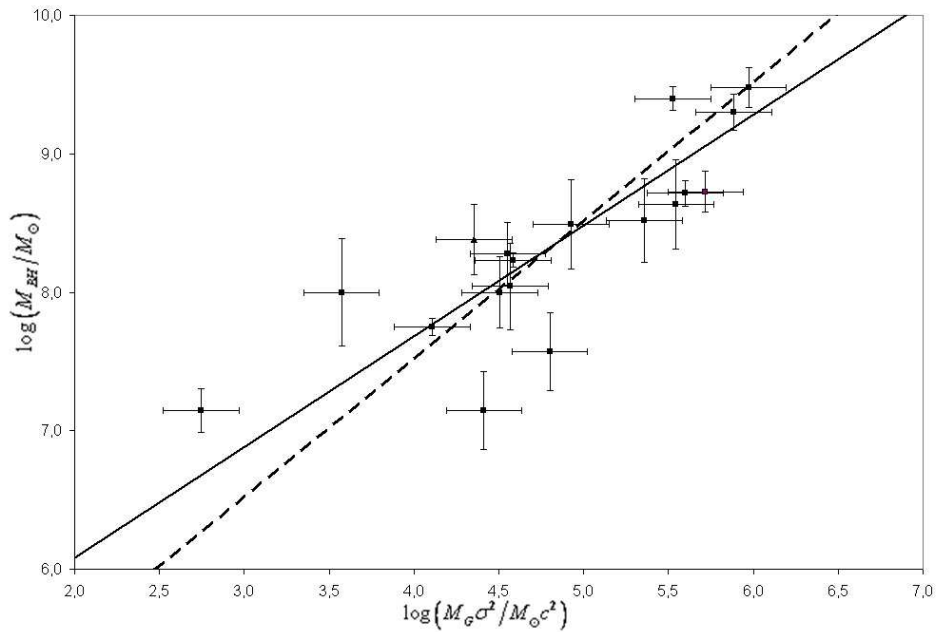}}
\caption{SMBH mass versus $M_G \sigma^2/c^2$ of the host
elliptical galaxy. The solid line is the best fit (11) with Fitexy
Routine, the dashed line is the best linear fit ($m = 1$, $b= 3.55
\pm 0.07$) with the exact formulas of the least squares method
(see appendix, eqs. A3 and A4). The error bars for $M_{BH}$ are
from \cite{tre} and we assume $\Delta LogM_G =0.18$ and for
$\sigma$ a relative error of $5\%$} \label{Fig. 1}
\end{figure*}

\begin{figure*}
\centering \resizebox{\hsize}{!}{\includegraphics{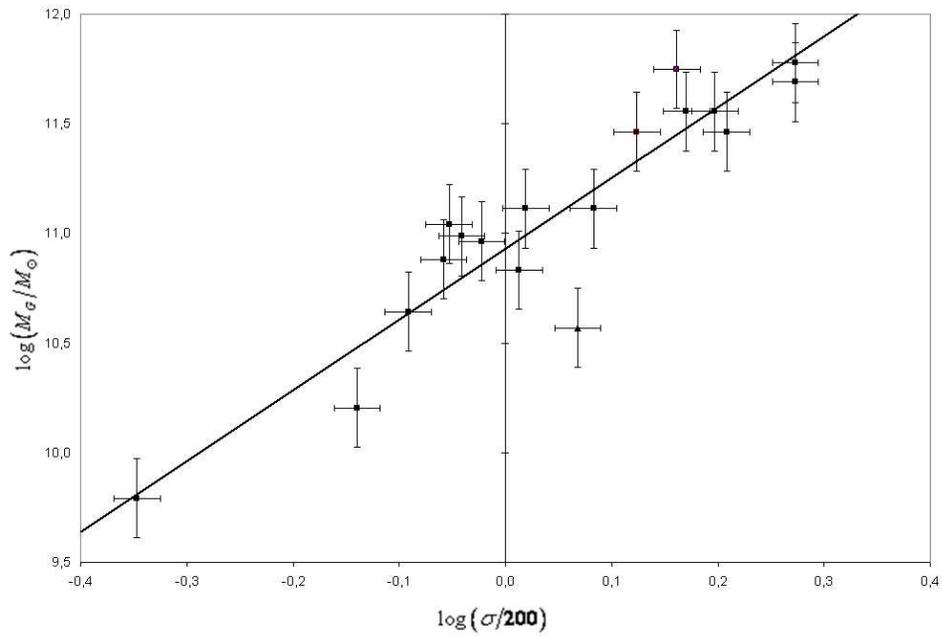}}
\caption{The mass of the galaxy $M_G$ versus $\sigma/200$ of the
host elliptical galaxy. The solid line is the best fit (14) with
Fitexy Routine. The error bars are such that $\Delta LogM_G =0.18$
and $\sigma$ has a relative error of $5\%$. We denote with a
triangle the galaxy NGC5845 with the largest residual.}
\label{Fig. 2}
\end{figure*}

\begin{figure*}
\centering \resizebox{\hsize}{!}{\includegraphics{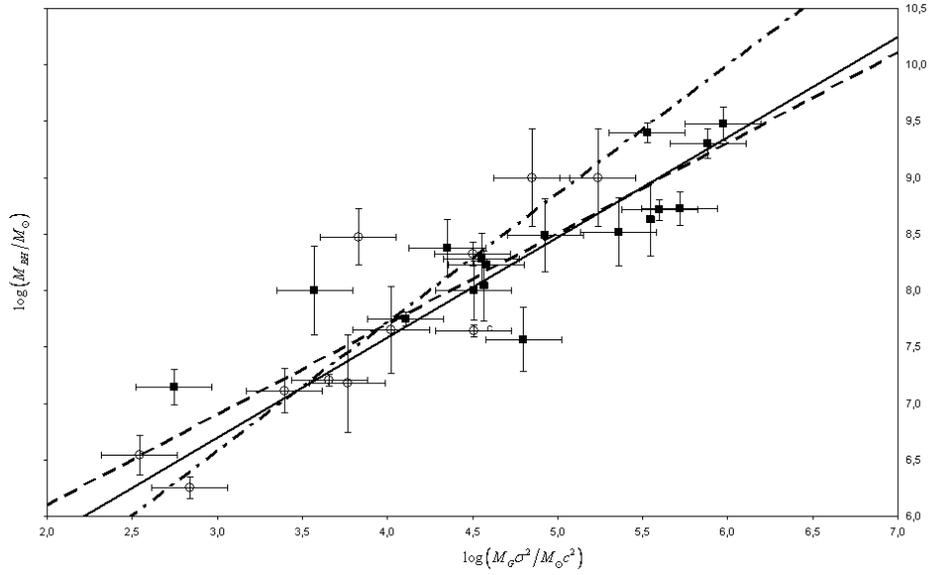}}
\caption{SMBH mass versus $M_G \sigma^2/c^2$ of the host galaxy.
The dashed line is the best fit (11) with Fitexy Routine for the
18 ellipticals denoted with boxes, the dot - dashed line is the
best fit (15) for the remaining 11 lenticular and spiral galaxies
denoted with circles. Finally the solid line is the best fit (16)
for the whole sample D of 29 galaxies. The error bars for $M_{BH}$
are from \cite{tre} and we assume $\Delta LogM_G =0.18$ and for
$\sigma$ a relative error of $5\%$} \label{Fig. 3}
\end{figure*}

\end{document}